\documentclass[nocopyright]{cimento}

\usepackage{graphicx,amsmath}  

\newcommand{\ud}{{\rm{d}}}

\title{Recent progress in modeling neutrino-nucleus interactions for oscillation experiments}
\author{Maria B.~Barbaro
}
\instlist{\inst{}
Dipartimento di Fisica, Universit\`a di Torino - Torino, Italy \\ 
 INFN, Sezione di Torino - Torino, Italy
}

\begin{document}

\maketitle

\begin{abstract}
The analysis of accelerator-based neutrino experiments crucially  depends on the good control of nuclear effects in the neutrino interactions with the detectors.
Recently the focus of both experimental measurements and theoretical studies has moved from the inclusive channel, where only the final lepton is detected, to the more challenging semi-inclusive reactions, where one or more hadrons are detected in coincidence with the lepton. 
In this contribution we present our recent research of the semi-inclusive $(\nu_\mu,\mu p)$ reaction on carbon for the quasielastic and the two-particle-two-hole channels.  We show some representative results and their comparison with experimental data, illustrating the robustness of the theoretical approach and discussing possible improvements to the model.
\end{abstract}

\section{Introduction}

The interpretation of neutrino oscillation data strongly relies on the accurate description of neutrino–nucleus cross sections in the GeV energy domain. Beyond increasing the precision of 
the neutrino mixing angles, current (T2K and NOVA) and next-generation (HyperK and DUNE)  
experiments aim to probe the leptonic CP-violating phase, crucial for understanding the matter–antimatter imbalance of the Universe, determine the neutrino mass hierarchy, and explore  physics beyond the Standard Model. 
Because detectors employ heavy nuclei (carbon, oxygen, argon), systematic uncertainties are dominated by nuclear effects.
Achieving the physics goals of future facilities will therefore depend on minimizing  these 
uncertainties across a broad kinematic range: from quasielastic (QE) scattering, where a neutrino interacts with a single bound nucleon, to deep inelastic scattering (DIS), which probes the  quark structure, covering the full energy spectrum across resonance excitation and multinucleon emission~\cite{NuSTEC:2017hzk}.

The extraction of the oscillation parameters from data depends directly on how well the neutrino energy $E_\nu$, which is distributed according to a broad flux in the incident beam,  is reconstructed from the kinematics of the observed particles: this leads to overlapping contributions from different processes in the same experimental signal, usually classified by the detected particles. For example, some events are labeled “QE-like” or “QE$0\pi$”, indicating that no pions are detected; however, true QE events cannot be isolated, since non-QE processes—such as two-nucleon emission (“2p2h” or two-particle–two-hole excitations)—also contribute significantly to the QE-like signal.

Until recently, most studies have focused on 
inclusive reactions, where only the outgoing lepton (a muon or an electron) is detected.
While calculations of this cross section in the QE$0\pi$ channel still show some spread, the current precision of inclusive data  
is insufficient to discriminate among them, as many details of nuclear dynamics are averaged out by integration over the hadronic variables. 
More recent efforts have shifted toward measuring and predicting semi-inclusive cross sections, where the scattered lepton is detected in coincidence with at least one ejected nucleon. These measurements are significantly more sensitive to nuclear effects than inclusive ones and can place stronger constraints on theoretical models.

In this context, here we briefly present  the most recent developments of our research in modeling the quasi-elastic and the 2p2h contributions to the semi-inclusive charged-current (CC) neutrino cross section $(\nu_\mu,\mu p)$.  More details on these calculations can be found in Refs.~\cite{Franco-Patino:2022tvv,Franco-Patino:2023msk,Belocchi:2024rfp,Belocchi:2025eix}.

 \section{Semi-inclusive CCQE scattering}

We consider an incoming neutrino of momentum $\mathbf{k}$ scattering off a nucleus $A$. In the final state a lepton with momentum $\mathbf{k'}$ and a nucleon with momentum $\mathbf{p_N}$ are detected in coincidence. 
Assuming that the impulse approximation (IA) is valid, {\it i.e.}, the  neutrino interacts only with one of the bound nucleons, the semi-inclusive 
neutrino-nucleus cross section averaged over the neutrino flux $\Phi(k)$ reads
\begin{equation}
\label{semi-inclusive}
		\langle \frac{d\sigma}{d\mathbf{k'} d\mathbf{p_N}}\rangle=\frac{G_F^2\cos^2{\theta_c}k'^2p_N^2}{64\pi^5}\int dk\frac{W_B}{E_Bf_{\text{rec}}}L_{\mu\nu}W^{\mu\nu} \, \Phi(k) ,
\end{equation}	
where $G_F$ is the Fermi constant, $\theta_c$ is the Cabibbo angle, $W_B$ and  $E_B$ are the invariant mass and energy of the residual nucleus $B$ (left, in general, in an excited state), $L_{\mu\nu}$ and $W^{\mu\nu}$ are the leptonic and hadronic tensors, and $f_{\text{rec}}$ is the recoil factor. 
	 	All information about the nuclear structure and final state interaction (FSI) effects is contained inside the hadronic tensor, constructed as the bilinear product of the matrix elements of the nuclear current operator $\hat J^\mu$ between the initial nuclear state $\left|A\right>$ and the final hadronic state $\left|B,p_N\right>$.

    We describe the initial nuclear state as a product of relativistic mean field (RMF) single-particle states, labeled by an index $\kappa$. The associated single-shell hadron tensor is
\begin{equation}\label{eq:hadronic_tensor}
		W_\kappa^{\mu\nu} = \rho_\kappa\left(E_m\right)\sum_{m_j, s_N}J_{\kappa, m_j, s_N}^{\mu}J_{\kappa, m_j, s_N}^{\nu*}\,,
        \end{equation}
where  $\rho_\kappa(E_m)$ is the missing energy density, with $E_m = W_B + m_N - M_A$, and  current matrix element is
	\begin{equation}
	\label{eq:current}
			J_{\kappa, m_j, s_N}^\mu = \int d\mathbf{r}\, e^{i\mathbf{r}\cdot\mathbf{q}}\overline{\Psi}^{s_N}\left(\mathbf{p_N},\mathbf{r}\right)\left(F_1\gamma^\mu +\frac{iF_2}{2m_N}\sigma^{\mu\nu}Q_\nu \right.
			 \left. + G_A\gamma^\mu\gamma^5 + \frac{G_P}{2m_N}Q^\mu\gamma^5\right) \Phi_\kappa^{m_j}\left(\mathbf{r}\right)\, ,
	\end{equation}
 $m_j$ being the third component of the total angular momentum $j$ of the bound nucleon and $s_N$ the spin projection of the ejected nucleon.
	The bound nucleon wave function  $\Phi_\kappa^{m_j}$ is obtained by solving the Dirac equation in coordinate space in presence of two scalar and vector potentials, $S(r)$ and $V(r)$, fitted to the nuclear ground state properties. 
	Effects beyond the pure shell model approach are introduced  through a depletion of the occupation of the shell model states, and the inclusion of high missing energy nucleons originating from long- and short-range correlations in the initial state. 
  The missing energy density  for $^{12}$C used in this work is fitted to reproduce the missing energy profile obtained from the spectral function calculation of Ref.~\cite{Benhar:1994hw}. 
	The scattered nucleon, described by the wave function ${\Psi}^{s_N}$, experiences final state interactions with the residual nucleus. We have considered two different treatments of FSI:
1) the energy-dependent RMF (ED-RMF) model, where the ejected nucleon  is a scattering solution of the Dirac equation with the same RMF potential used to describe the bound nucleon, multiplied by a phenomenological function that weakens the potential for increasing nucleon momenta, in order to ensure that the PWIA is recovered at very high energy, and 2) 
the Relativistic Optical Potential (ROP) approach, 
where the ejected nucleon moves across the residual hadronic system under the influence of a complex  relativistic optical potential fitted to reproduce elastic proton-nucleus scattering data. We consider as a reference also 
the Relativistic Plane-Wave IA (RPWIA), in order to assess the importance of FSI. 

\begin{figure}[!htbp]
		\centering
\includegraphics[width=0.7\textwidth]{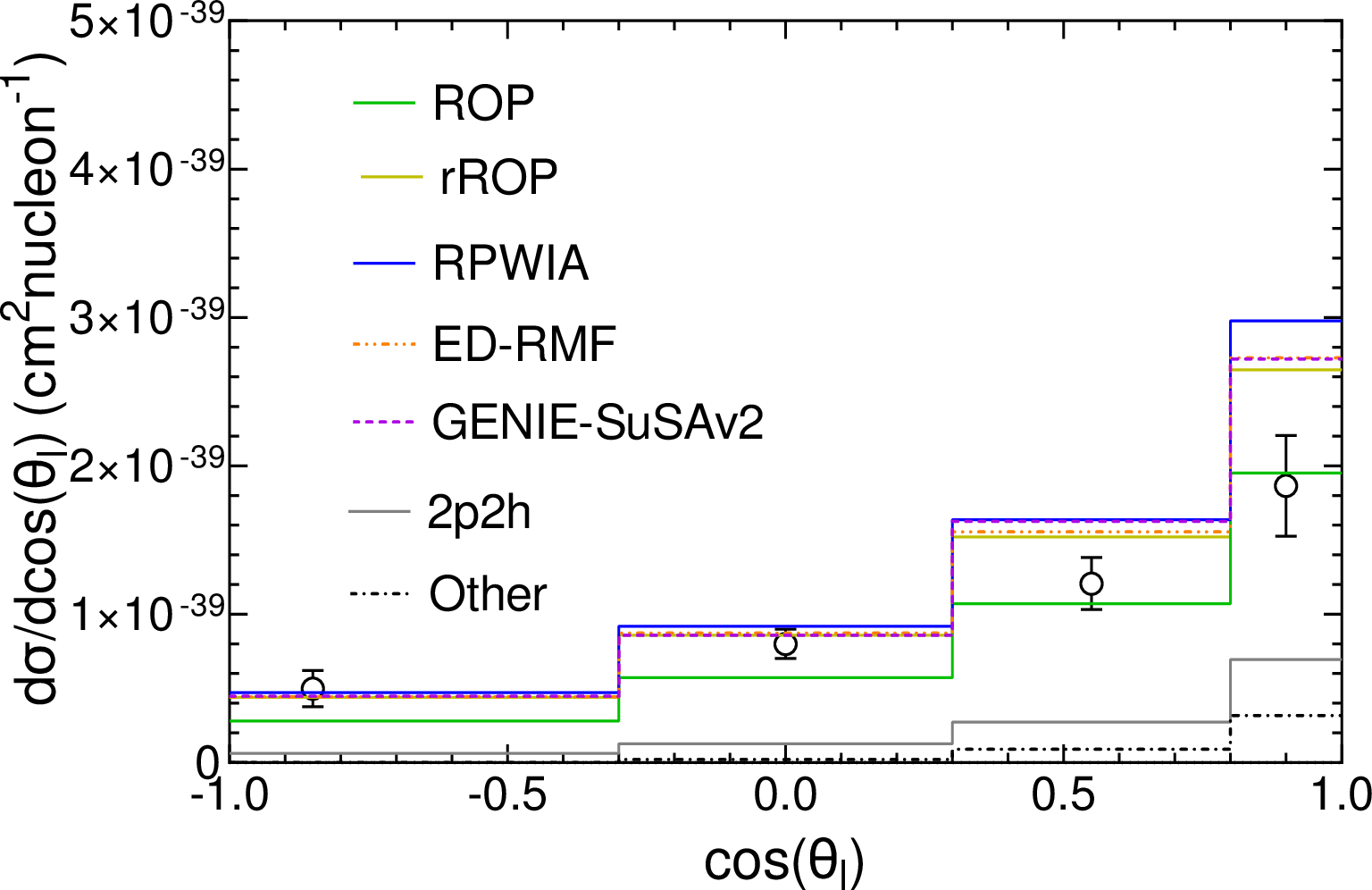}
	\caption{\label{fig:T2K_angle_semi-inclusive} T2K CC0$\pi$ semi-inclusive $\nu_\mu-^{12}$C cross section with protons in the final state with momenta above 0.5 GeV as function of the muon scattering angle. All curves include the 2p2h and pion absorption contributions (also shown separately), evaluated using GENIE. Figure adapted from Ref.~\cite{Franco-Patino:2022tvv}.}
	\end{figure}

Several comparisons of the prediction of this calculation with T2K, MINERvA and MicroBooNE data can be found in Refs.\cite{Franco-Patino:2022tvv} and \cite{Franco-Patino:2023msk}. A representative example is shown in 
Fig.~\ref{fig:T2K_angle_semi-inclusive}, where different models are compared with the T2K semi-inclusive CC0$\pi$Np cross-section data~\cite{T2K:2018rnz}. 
 Beyond the models illustrated above, in this figure we also show the "rROP" curve, corresponding to a purely real optical potential, which would be appropriate to describe an inclusive process, and the "GENIE-SuSAv2" results, which corresponds to the implementation in the GENIE generator of the SuSAv2 model~\cite{Amaro:2004bs,Gonzalez-Jimenez:2014eqa}. The latter is based on the connection between electron- and neutrino-nucleus scattering and on the scaling properties of inclusive $(e,e')$ data in the QE regime, and implements nuclear dynamics through inclusive scaling functions evaluated in the RMF model.
 Moreover, all the curves shown in Fig.~\ref{fig:T2K_angle_semi-inclusive} include a 2p2h contribution, representing the cross section associated to emission of two nucleons. Similarly to the GENIE-SuSAv2 curve, this contribution is evaluated on the basis of the {\it inclusive} calculation of Ref.~\cite{RuizSimo:2016rtu} 
 and  its application to semi-inclusive data requires strong assumptions which go beyond  the original model.

We can observe in Fig.~\ref{fig:T2K_angle_semi-inclusive} that FSI have a very strong impact on the cross section and tend to suppress it. At a quantitative level, this effect varies depending on the prescription used to describe FSI. The ROP model provides the strongest suppression and the best overall description of the data,
although for backward angles
it underestimates the cross section. 
However, these conclusions are strongly dependent on the GENIE simulation of the 2p2h contribution, which is sizeable - especially at forward angles - and is affected by the above mentioned problem: it is based on an inclusive calculation and its extrapolation to semi-inclusive data is questionable.
In order to draw more reliable conclusions one should perform a microscopic calculation of this contribution, as illustrated in the next paragraph.
    
    \section{2p2h contribution to semi-inclusive neutrino scattering}

Electron  scattering studies have shown long ago that  two-nucleon knockout  is the key process occurring in the kinematical region between the QE and the $\Delta$-resonance peaks.
The importance of this process in neutrino studies was first recognized when the CC$0\pi$ MiniBooNE data of neutrino-carbon cross sections~\cite{MiniBooNE:2010bsu} revealed a big discrepancy with the RFG prediction. Theoretical work incorporating 2p2h 
explained the discrepancy, and this channel is now widely accepted as a crucial contribution to the cross section. 
Whereas several calculations exist for the 2p2h inclusive cross section, the semi-inclusive channel remains almost unexplored.
 To fill this gap, we have developed a fully microscopic model for semi-inclusive 2p2h neutrino-nucleus interactions, extending the electromagnetic $(e,e'p)$ framework of Ref.~\cite{Belocchi:2024rfp} to the weak reaction $(\nu_\mu,\mu p)$.  Our approach is briefly outlined below.

The  two-nucleon knockout process is mainly driven by two-body currents, that we choose to be meson-exchange currents (MEC) in which a pion is exchanged between two nucleons, with the possible excitation of a virtual $\Delta$ resonance. 
The full expression of the current can be found in Ref.~\cite{Belocchi:2025eix}.
The associated 2p2h hadronic tensor is evaluated in the 
framework of the RFG, a system of free nucleons described by Dirac spinors, correlated only by the Pauli exclusion principle. 
At fixed  energy of the incident neutrino\footnote{To compare with experimental data, a folding integral over the neutrino flux must be performed.} the 2p2h hadron tensor  reads~\cite{Belocchi:2025eix}
\begin{eqnarray}
  W_{\rm 2p2h }^{\mu \nu}&= V(2\pi)^3\frac{m_N^2}{(2\pi)^3E_{p_1}}
  \theta(|\mathbf{p_1}|-p_F|)
 \int
 \frac{m_N \, \ud \mathbf{h_1}}{(2\pi)^3E_{h_1} }\theta(p_F-|\mathbf{h_1}|)
  \int
  \frac{m_N \, \ud \mathbf{h_2}}{(2\pi)^3E_{h_2} }\theta(p_F-|\mathbf {h_2}|)
  \nonumber
\\  &\times \int \frac{m_N\, \ud \mathbf{p_2}}{(2\pi)^3E_{p_2}}  \, w_{\rm 2p2h }^{\mu \nu}(h_1,h_2,p_1,p_2)\, \delta^4(q +h_1+h_2 -p_1-p_2)\theta(|\mathbf{p_1}|-p_F|)\,,
  \label{eq:WmunuN}
  \end{eqnarray} 
  where the nucleon detected has been chosen, without loss of generality, to be the one with momentum \(\mathbf{p_1}\), 
  $V$  is the volume of the system, Pauli blocking is encoded in the two step functions acting on the final-particle momenta $\mathbf{p_1}$ and $\mathbf{p_2}$, and $w^{\mu\nu}_{2p2h }$ is
the elementary tensor
obtained from the matrix elements of 
the MEC operator.
To make the RFG model more realistic, an energy shift $E_{\rm shift}$=40 MeV is introduced  to account for the nucleon binding energy and some FSI effects. This model has  been validated in Ref.~\cite{Belocchi:2024rfp}, where it has been shown to provide a good description electron scattering of $(e,e'p)$ data in kinematical conditions where the 2p2h contribution is dominant. 

\begin{figure}
    \centering
    \includegraphics[width=0.27\linewidth,angle=270]{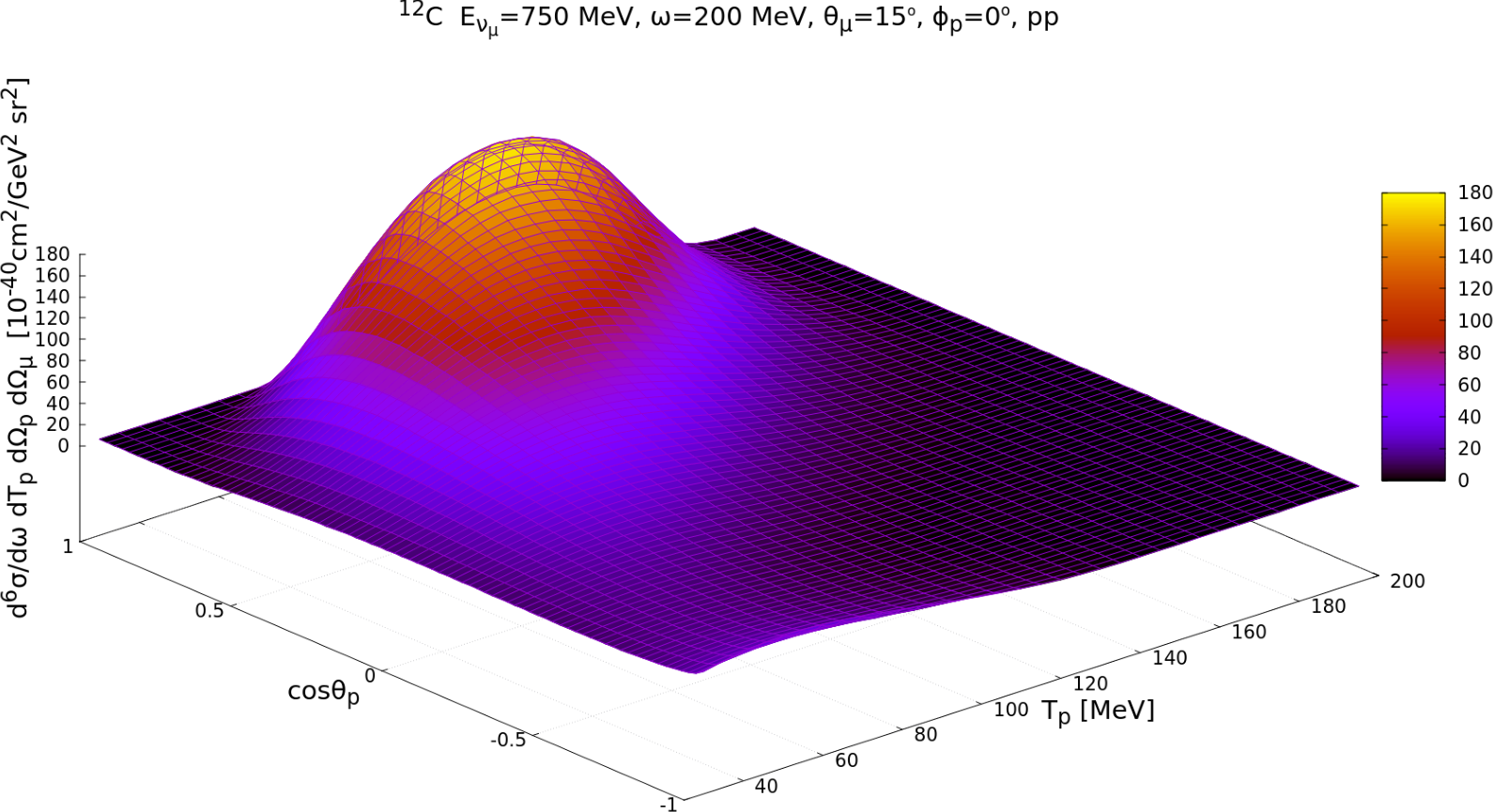} \includegraphics[width=0.27\linewidth,angle=270]{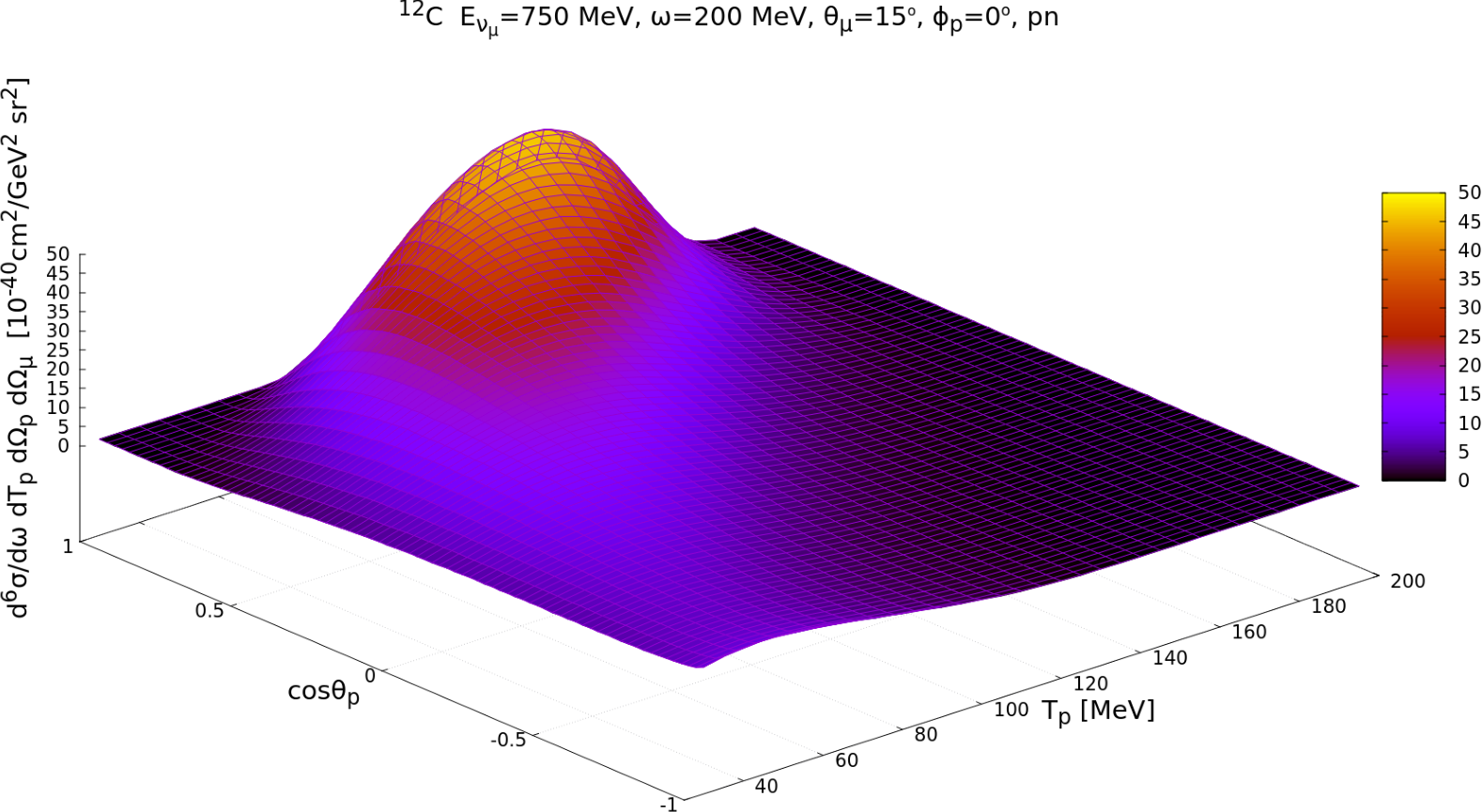}
    \caption{The $\nu_\mu-^{12}$C 2p2h, computed at energy \(E_{ \nu_\mu}= 750\) MeV and transferred energy \(\omega= 200\) MeV, scattering angle \(\theta_\mu =15\)° and azimuthal final proton angle \(\phi_p= 0\)°, is displayed versus the polar angle \(\theta_p\) and kinetic energy \(T_p\) of the final proton. The two plots correspond to the contributions of the  \(pp\) (left) and \(pn\) (right) final states. Figure adapted from Ref.~\cite{Belocchi:2025eix}.}
    \label{fig:3D-15pp_pn}
\end{figure}

In Fig.\ref{fig:3D-15pp_pn} the $(\nu_\mu,\mu p)$ cross section on carbon, computed at fixed neutrino energy,
is displayed as a function of the proton polar angle and kinetic energy, separating the cases of $pp$ and $pn$ emission.
The cross section displays a well-defined peak nearby \(\theta_p=0^\circ\), the so-called ``parallel kinematics''. 
We also observe that the \(pp\) channel exhibits a broader bump than the \(pn\) channel, which is more localized and has a peak at higher \(T_p\) values. The results clearly show that $pp$ emission dominates over $pn$ emission, with a the ratio $pp/pn$ of \(\simeq 4\) for this kinematics.

To make contact with the existing experimental data, we have also computed the flux-folded differential semi-inclusive cross sections, using the T2K semi-inclusive  measurement of Ref.~\cite{T2K:2018rnz} as a reference and using the same kinematical cuts, which include restrictions on the outgoing muon and proton kinematics.
\begin{figure}
    \centering
\includegraphics[width=0.4\linewidth,angle=270]{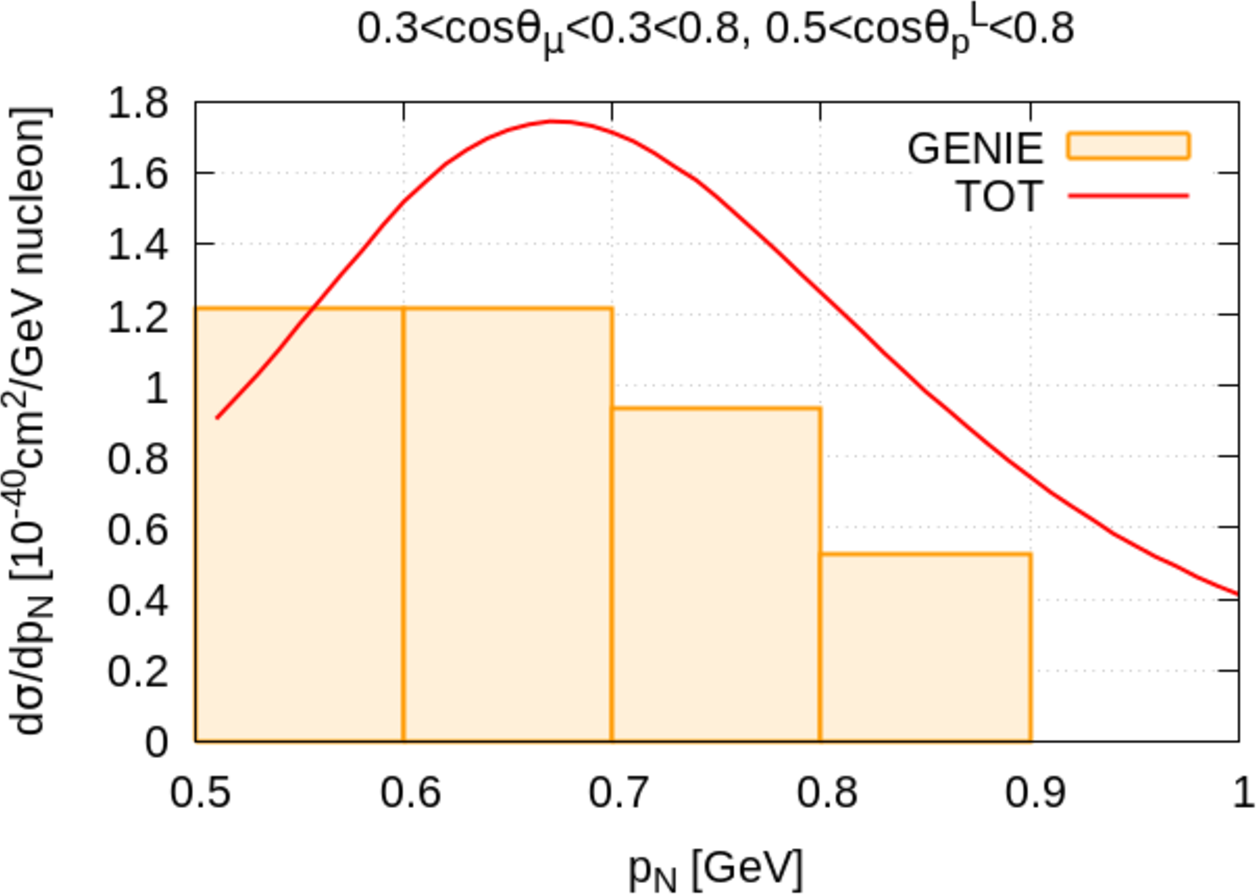}\\
    \caption{2p2h contribution to the flux-averaged  \(1\mu\)CC\(0\pi\) Np cross section, evaluated using the T2K flux  and kinematic cuts of Ref.~\cite{T2K:2018rnz}, displayed versus the proton  momentum and compared with the GENIE simulation~\cite{Dolan:2019bxf}.     Figure adapted from Ref.~\cite{Belocchi:2025eix}.}
    \label{fig:EW-Genie_comp}
\end{figure}
In Fig.~\ref{fig:EW-Genie_comp} the flux-integrated  2p2h cross section 
is displayed as a function of the leading ejected proton momentum (the leading proton is the one having the highest momentum in the pair).
Our result is compared with the predictions 
of the GENIE implementation of the SuSAv2-MEC model~\cite{Dolan:2019bxf}. In spite of the general agreement in terms of shape and order of magnitude, systematic discrepancies are observed bewteen the two results: our curve shows a higher strength compared to the GENIE simulation, with a peak occurring at a higher final leading proton momentum. This discrepancy  is not unexpected, since the two approaches are  quite different: the  GENIE result is based on an inclusive theoretical model and requires certain assumptions in order to ``extract'' semi-inclusive predictions from the inclusive results, while our calculation deals correctly with both the final lepton and hadron  kinematic variables.
On the other hand, the  GENIE result benefits from the incorporation of additional nuclear effects, absent in our calculation, such as FSI, which are modeled using the semi-classical cascade approach.  In general, FSI tend to broaden and shift the differential cross section toward lower final proton energy and momentum. These effects must be included in our calculation to perform a more consistent comparison.

\section{Conclusion}

We have outlined our most recent studies of neutrino-nucleus interactions, focusing on quasielastic and 2p2h reactions in terms of both the final lepton and hadron kinematics. Selected results have been shown and compared with T2K  data and  Monte Carlo simulations.
The analysis indicates that future work should focus on the inclusion of initial and final state interactions in the 2p2h calculation, currently performed using a RFG framework. Comparison with forthcoming data will also help  better constrain the model for FSI in the quasielastic channel, which remains affected by significant uncertainties.
Additionally, there is growing interest in the inelastic domain, which represents a crucial ingredient of the high-energy cross sections relevant for the future DUNE experiment. We have recently addressed this problem in Ref.~\cite{Gonzalez-Rosa:2024udj}.


\end{document}